\documentstyle[12pt,twoside,fleqn,espcrc1,epsfig]{article}  
\begin{document}  
\def\J{$J/\psi$}
\def\j{J/\psi}
\def\P{$\psi'$}
\def\p{\psi'}
\def\U{$\Upsilon$}
\def\u{\Upsilon}
\def\C{{\bar c}c}
\def\b{{\bar b}b}
\def\q{{\bar q}q}
\def\Q{{\bar Q}Q}
\def\F{$\Phi$}
\def\f{\Phi}
\def\T{$\hat t$}
\def\d{$d{\hat{\sigma}}$}
\def\t{\tau}
\def\E{\epsilon}
\def\a{\alpha_s}
\def\d{\lambda}
\def\la{\Lambda_{\rm QCD}}
\def\T{$T_f$}
\def\n{$n_b$}

\def\L{\cal L}

\def\l{\lambda}

\def\e{\epsilon}

\def\lsim{\raise0.3ex\hbox{$<$\kern-0.75em\raise-1.1ex\hbox{$\sim$}}}
\def\gsim{\raise0.3ex\hbox{$>$\kern-0.75em\raise-1.1ex\hbox{$\sim$}}}

%
%DEFINE JOURNAL NAMES
\def\CMP{{ Comm.\ Math.\ Phys.\ }}
\def\NP{{ Nucl.\ Phys.\ }}
\def\PL{{ Phys.\ Lett.\ }}
\def\PR{{ Phys.\ Rev.\ }}
\def\PRep{{ Phys.\ Rep.\ }}
\def\PRL{{ Phys.\ Rev.\ Lett.\ }}
\def\RMP{{ Rev.\ Mod.\ Phys.\ }}
\def\ZP{{ Z.\ Phys.\ }}

\def\etal{{\sl et al.}}
  
\input{epsf.tex}  
\newcommand{\be}{\begin{eqnarray}}  
\newcommand{\ee}{\end{eqnarray}}  
\newcommand{\ben}{\begin{eqnarray*}}  
\newcommand{\een}{\end{eqnarray*}}  
\title{Theoretical Interpretations of $J/\psi$ Suppression:\\  
A Summary\footnote{Invited plenary talk presented at Quark Matter'97 Conference, Tsukuba, Japan}}  
\vskip0.1cm  
%\author{D. Kharzeev}  
\maketitle  
\begin{flushleft}  
D. Kharzeev\\  
\vskip0.3cm  
RIKEN-BNL Research Center, Brookhaven National Laboratory, Upton NY 11973, USA  
\end{flushleft}  
\vskip0.3cm  
\begin{abstract}  
The strong ``anomalous" \J~ suppression observed recently by the   
NA50 Collaboration at CERN SPS has attracted considerable attention.   
Is it the first signature of a long-awaited quark-gluon plasma,   
or just a peculiar combination of ``conventional" effects acting together   
to produce the puzzling pattern observed experimentally?   
This talk is an attempt to summarize the theoretical explanations   
proposed during the last two years.   
  
\end{abstract}  
  
\section{INTRODUCTION}  
  
The data on \J\ production in high energy nuclear collisions \cite{Ramello}   
have never been more puzzling,  
the debates on the origin of \J\ suppression have never been   
more heated, and the topic is rapidly attracting attention of   
physicists, including even those who have never been interested 
in it before.   
What is at the heart of the debate, and why is \J\ suppression   
so interesting?       
  
\J\ is a peculiar object. Since it is made of heavy quark and antiquark,   
its size is small enough to make perturbation theory meaningful,   
and the description of \J\ properties and decays was among the first   
succesful applications of perturbative QCD.   
It is not sufficiently small though to make non-perturbative effects   
totally negligible; however, they can be analyzed in a systematic way,   
providing a unique information about the strength of soft gluon fields in   
QCD vacuum \cite{SVZ}. The \J\ thus serves a special r\^{o}le 
of the ``borderguard"   
\cite{Dokshitzer} on the mysterious border of perturbative world of   
quarks and gluons and non-perturbative world of hadrons.   
When the structure of the vacuum is violently disturbed in a high   
energy nuclear collision and the soft and colorless hadronic world   
suffers the intrusion of abundantly produced colored gluons and quarks,   
these borderguards are among the first to suffer. What is even more   
important, \J's, unlike light hadrons, cannot be easily reproduced at   
later stages of the collision, and their disappearance is documented in   
the dilepton spectra. These considerations laid the basis for the   
proposal \cite{MS} to use \J's and other heavy quarkonium states   
for the diagnostics of hot and dense QCD matter.   
  
Observed experimentally by the NA38 Collaboration in 1987,   
\J\ suppression therefore excited considerable interest, and immediately   
triggered debates as to the origin of the effect.   
It has become clear eventually that a ``conventional" approach   
can explain all features of the \J\ suppression observed in   
nuclear collisions with light ion projectiles.  
However the controversy resumed when the new Pb-Pb results of the NA50 
Collaboration \cite{Ramello} have been revealed.

\section{IS THE OBSERVED \J\ SUPPRESSION ANOMALOUS?}  
  
Before we start discussing various theoretical explanations of the   
observed phenomenon, let us recall briefly what is actually   
observed and why it is so surprising. The entire set of the   
\J\ production data from $pA$ and $AB$ \cite{NA38} collisions 
available before the advent of the Pb beam at 
CERN SPS has been found \cite{KLNS} to be 
consistent   
with the nuclear absorption model \cite{GH}. However the fact that  
$\psi'/J/\psi$ ratio in $pA$ collisions does not depend    
on the atomic number $A$ (see \cite{Carlos} for a comprehensive 
compilation of the available data) 
shows that one cannot interpret   
the observed nuclear attenuation as the result of the absorption   
of physical \J\ and $\psi'$ states in nuclear matter: $\psi'$ is known   
to have a radius about twice larger than that of \J\, and is   
expected to be absorbed with a much larger cross section.   
Another argument against the na{\"{\i}}ve picture of \J\ absorption  
in nuclear matter    
stems from the magnitude of the extracted cross section, which   
appears to be approximately two times larger than the \J\   
absorption cross section extracted from the data on \J\ photoproduction   
on nuclear targets at small energies \cite{photo} and from 
the VMD analyses of   
photoproduction on protons. It seems quite safe therefore to state that   
the gross features of the data on \J\ production on nuclear targets   
available before the new Pb-Pb results are consistent with   
pre-resonance absorption. (This concerns the integrated cross  
sections, which are determined mainly by the central region; we leave  
aside for the moment the interesting question of $x_F$, or rapidity,  
dependence of the \J\ suppression).   
  
A model which naturally accomodates the listed above features of the   
\J\ production on nuclear targets \cite{KS2}  
is motivated by the presence of the higher Fock states in the \J\ wave   
function \cite{NR}, revealed by the recent Tevatron results \cite{TeV}.  
At small $p_T$, the \J\ production is assumed to proceed through   
the formation of the color singlet pre-resonance $|\bar{c}c g\rangle$ state.   
Even though the proper formation time of the physical \J\ and $\psi'$ states  
is estimated to be rather short, about $0.3$ fm,  
the pre-resonance state can propagate through the entire volume of  
the nucleus already at SPS energies due to the Lorentz  
dilatation factor. The target independence of the $\psi'/J/\psi$ ratio  
in $pA$ collisions is natural in this picture.   
  Furthermore, since the $\bar{c}c$ pair is produced at short  
distances $\sim 1/ 2 m_c$,   
much smaller than the inverse transverse momentum of the collinear gluon,    
the color structure of this state is that of a color dipole formed by two   
octet charges - the gluon and the almost pointlike $(\bar{c}c)_8$.   
The interaction of such a dipole with external color fields is   
enhanced, compared to the usual triplet-antitriplet dipole structure of  
the \J\,   
 by the color Casimir factor of $9/4$; one therefore expects an accordingly   
larger absorption cross section for this pre-resonance state.   
It should be noted however that a first-principle QCD calculation based  
on this picture, which would allow to promote the model to a consistent  
theoretical approach, is still lacking at present \cite{Mueller}.       
  
The new Pb-Pb data in the peripheral region of $E_T 
\leq 50$ GeV are consistent, within error bars and uncertainty in the 
value of the absorption cross section, 
with the pre-resonance absorption calculated in Glauber theory with the 
cross section of $\sigma_{abs}=7.3\pm 0.6$ mb, extracted from the previous 
\J\ production data \cite{KLNS} (see Fig. 1, which shows the result of 
Glauber calculation 
with the central value of $\sigma_{abs}=7.3$ mb)\footnote{The peripheral 
Pb-Pb data alone suggest a somewhat larger value of $\sigma_{abs}$, 
but it is consistent within the error bars with $\sigma_{abs}=7.3\pm 0.6$ mb 
extracted from the previous data.}.

%------------------------------------------------------------------
%%                   figure 1
\begin{center}\mbox{
\epsfig{file=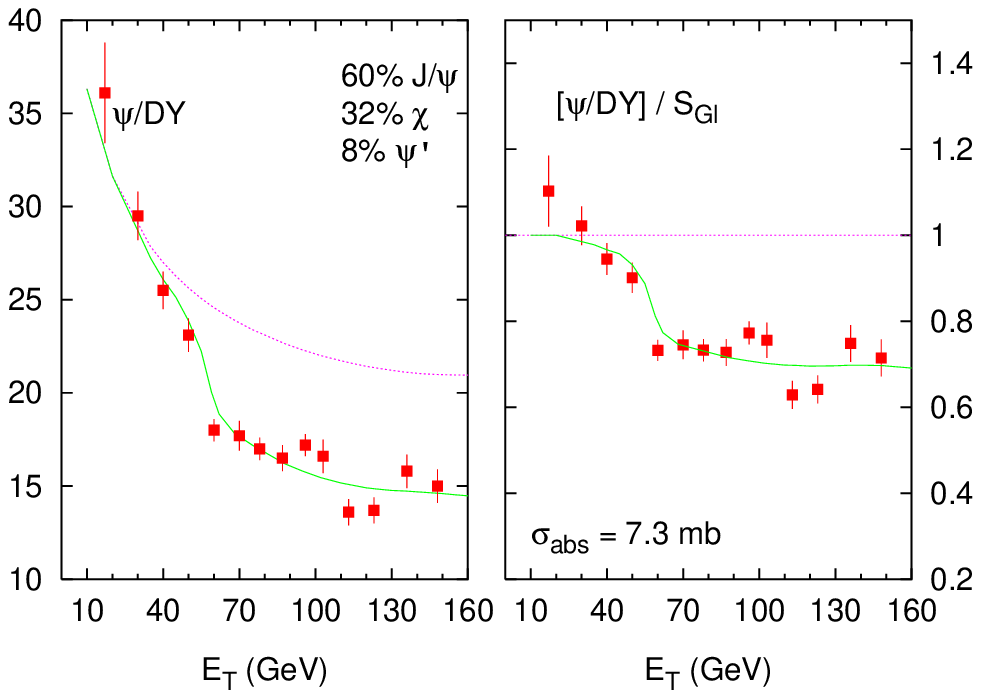,height=10cm}}\end{center}
\begin{center}Figure 1. Left: the \J/DY ratio in Pb-Pb collisions 
\cite{Ramello} versus the prediction of the pre-resonance absorption 
model \cite{KLNS} with $\sigma_{abs}=7.3\ {\rm mb}$ (the upper curve). 
Right: the same, normalized to the prediction of the pre-resonance 
absorption model. The lower curve is the result of the 
calculation based on the model of ref \cite{KNS} (see text).   
\end{center}

However around   
$E_T \simeq 50$ GeV (corresponding to the average impact parameter of  
$b \simeq 8 \pm 1$ fm) the \J/DY ratio jumps down, and deviates   
significantly from the predictions of Glauber model. The $\psi'$/DY   
ratio at the same time does not seem to show any discontinuities.   
Remarkably, the integrated Drell-Yan production cross section was    
found to be consistent with the $A \cdot B$ scaling established   
previously. 

To summarize, the \J\ suppression observed in Pb-Pb   
can indeed be considered ``anomalous'' -- it is    
different from what was observed before in the entire set   
of the \J\ production data accumulated prior to the NA50 experiment. 
We now proceed to the discussion of   
various theoretical explanations of this effect.            
  
\section{INITIAL STATE INTERACTIONS}  
  
Several authors \cite{init} have considered the effect of   
nucleon energy loss in the initial state  
on the \J\ production. Their idea can be briefly summarized as follows:   
in a nucleus-nucleus collision, the colliding nucleons loose energy   
before they produce \J's. Since at SPS we are still in the energy   
range where the \J\ production cross section is a steep function   
of the incident nucleon's momentum (see e.g. \cite{gavai}), 
this initial state energy loss   
will lead to a strong suppression of the \J\ production.   
The Glauber -- like approaches do not consider the energy loss mechanism   
and are therefore misleading, significantly underestimating   
the \J\ suppression expected from conventional mechanisms.   
At first glance, the argument looks correct, and seems to be well   
supported experimentally -- the effects of the nucleon energy loss   
in $pA$ and $AB$ collisions are well established \cite{busza}.   
The problem arises, however, when one recalls that the Drell-Yan   
pair production cross section, also measured in Pb-Pb collisions   
by the NA50, follows the $A \cdot B$ scaling law\footnote{The $A \cdot B$  
scaling of the cross section implies that {\it all}  
individual nucleon--nucleon collisions  
are equally effective in producing Drell-Yan pairs.} established previously   
in $pA$ and $SU$ data. Indeed, the high-mass ($M \geq 4$ GeV)   
Drell-Yan pair production cross section is also a steep function   
of the incident momentum at SPS energies (see e.g. \cite{gavin}), 
and if the initial state   
nucleon energy loss effects are important, one inevitably   
arrives to the conclusion that Drell-Yan pairs should also   
be strongly suppressed -- contrary to experimental observations.   
 It looks therefore that we have a difficulty reconciling the two   
well established experimental facts -- the existence of the nucleon   
energy loss   
in nuclear matter and the $A \cdot B$ scaling of the Drell-Yan   
pair production.   
Is this a paradox?  
 
The answer has been known for quite a   
long time \cite{Feinberg}, \cite{Gribov}: quantum mechanics implies that at high   
energies, soft processes develop over large longitudinal distances.   
Consider, for example, a proton with momentum $P$ which undergoes   
an inelastic diffractive interaction inside the nuclear target, transforming  
itself into a cluster of particles of invariant mass $M_*$.   
Let us consider the amplitude of this process in the momentum representation:   
\be  
M(q_L) = \int dz\ e^{i q_L z}\ M(z), \label{amp}  
\ee  
where $q_L$ is the longitudinal momentum transfer (we have suppressed   
the transverse coordinate integration). Energy conservation implies that   
\be   
q_L = \sqrt{E^2-M_p^2} - \sqrt{E^2-M_*^2} \simeq \frac{M_*^2 - M_p^2} {2 P},   
\label{kin}  
\ee    
where $M_p$ is the mass of the proton.  
Because of the presence of oscillating exponential in   
(\ref{amp}),   
the most important contribution to $M(q_L)$ will come from   
the region where $q_L z \leq 1$, i.e. from the region with the longitudinal   
size of   
\be  
z \simeq \frac{2P} {M_*^2 - M^2} = \frac{P} {(M_* + M)/2}\ \cdot \  
\frac{1} {M_* - M}.  
\label{dist}  
\ee     
It is clear from (\ref{dist}) that inelastic interactions of the   
incident nucleon, responsible for its energy loss, at high energies   
develop over large longitudinal distances which grow proportionally to   
the initial momentum. The result (\ref{dist}) in a time-dependent   
picture can be interpreted as the product of proper formation time   
$\tau \simeq (M_* - M_p)^{-1}$ of the proton, given by the uncertainty  
relation, and the Lorentz factor $P / \bar{M}$,   
where $\bar{M} = (M_* + M_p)/2$ is the average mass of the wave  
packet consisting   
of a proton and the excited state with invariant mass $M_*$.    
(It is worth to note that we were able to deduce the existence of  
formation zone (\ref{dist}) starting from the mere assumption 
that the process  
is described by an {\it amplitude} (\ref{amp}), rather than  
a {\it probability}.) 
The data on the invariant mass distributions in the inelastic proton  
interactions   
show that in most collisions the invariant mass $M_*$ is not 
large \cite{Goulianos}; 
typically   
$M_* \leq 4$ GeV. The formula (\ref{dist}) implies therefore that  
already at SPS   
energies a typical inelastic collision develops over a distance comparable  
to,   
or larger than, the size of the nucleus. This means that the nucleons   
traversing the nucleus have not lost their energy yet; all of their   
incident momentum can still be utilised for a hard process, 
which is much better   
localized in the longitudinal direction. Quantum mechanics therefore   
provides a natural explanation of the apparent 
``paradox'', and forces us to discard  
the nucleon energy loss explanation of \J\ suppression.   
  
A different approach \cite{Rudy} considers initial state interactions on the   
parton level. In this way, formation time effects are implicitly taken   
into account: at high energies, because of the Lorentz factor,   
the time during which the nucleon traverses the nucleus is shorter  
than it takes   
for a signal to propagate through the nucleon's   
transverse size. This implies that different parton  
configurations of the nucleon   
will interact incoherently; one can therefore distinguish   
between the interactions of (anti)quarks and gluons from the incident   
nucleon. Because of the larger color charge, the gluons are expected   
to interact stonger than quarks inside the nucleus.  
Since the Drell-Yan pairs are   
produced (in the leading order in $\alpha_s$) by the quark-antiquark fusion,   
and the heavy quarks by the gluon-gluon fusion, one can try to  
reconcile the absense   
of initial state effects in Drell-Yan pair production with strong  
suppression        
observed for the \J\,even though the Drell-Yan data still do impose an   
important constraint on the model.     
 
Besides \J\ suppression, this mechanism should also cause suppression of  
the open charm production in $pA$ and $AB$ collisions.     
Even though the current data do not seem to show such suppression,  
I certainly agree with the authors of ref \cite{Rudy} that  
more data, particularly on correlated $\bar{D}D$  
production, are needed to clarify the issue. Nevertheless,  
before such data become available,  
let me present a theoretical argument\footnote{The arguments below  
are based on the discussion  
with Yu. Dokshitzer.} in favor of {\it universality} of quark and gluon  
depletion in nuclear matter at small $x$ (high energies and central  
region), which implies that the initial-state gluon absorption  
(or energy loss) is unlikely to be  
{\it the} mechanism responsible for the observed \J\ suppression.     
Indeed, the virtuality ordering in the QCD DGLAP \cite{DGLAP} evolution  
means that at small $x$, the heavy quarks and Drell-Yan pairs  
are generated  
{\it at the very end} of the parton ladder; the evolution at the  
preceding stages of the parton cascade is identical in both cases.  
Moreover, the gluons fusing to form a heavy quark pair, or quarks  
and antiquarks annihilating into a Drell-Yan pair, have a large  
virtuality and, at small $x$, small momentum -- therefore, they almost do not  
propagate inside the nucleus! To see how it works numerically, let us consider  
a nucleus--nucleus collision at $P = 200$ GeV of incident momentum.  
In the lab frame, the partons producing a heavy quark (or a Drell-Yan) pair  
of  
invariant mass $Q$ propagate the distance  
\be 
z \simeq \frac{P x}{Q} \ \ \frac{1}{Q}, \label{long} 
\ee 
where $x$ is the fraction of the incident nucleon's momentum carried  
by the parton before it splits into a $\bar{Q}Q$ or a Drell-Yan pair;  
the formula (\ref{long}) is, as usual, the product of the Lorentz factor and  
the proper formation time $1/Q$.  
In the central rapidity region of $y^* \simeq 0$ one has a simple  
kinematical relation  $x \simeq Q/\sqrt{s} \simeq Q/(2 M_p P)^{1/2}$. 
The value of $Q\simeq 3$ GeV leads to $x\simeq 0.15$;  
this is the region of $x$ where the sea partons begin to dominate  
over the valence quarks in the nucleon's wave function. Substituting  
this value of $x$ into (\ref{long}), we find that the final partons  
of DGLAP ladder propagate inside the nucleus the distance of less  
than $1$ fm. Since, as was stated above, the preceding part of the  
evolution is independent of the final stage, we do not expect  
any difference in the nuclear attenuation of quark-- and gluon--induced  
hard processes. In particular, the \J\ and Drell-Yan suppressions  
due to the initial state effects have to be the same.  
The situation will change, however, if we move out of the central region,  
since either $x_p$ or $x_t$ will then become large, involving  
the valence partons in the production process. 

To conclude this Section, let us note that independently  
of any dynamical details of initial state interactions, their effects are  
expected to increase {\it gradually} when the atomic number  
of the colliding nuclei grows. Therefore one certainly does not expect  
any discontinuities in \J/DY ratio arising from these effects.  
However initial state interactions are clearly interesting in their  
own right and should be studied in detail.   
 
\section{INTERACTION WITH HADRONIC SECONDARIES} 
 
The large number of hadronic secondaries in a typical nucleus-nucleus  
collisions naturally implies the possibility of final state interactions,  
absent in a $pp$ collision. In fact, one can even prove that  
final state interactions are important for charmonium  
production -- the $\psi'/J/\psi$ ratio in S-U collisions is known \cite{NA38} 
to drop  
by a factor of two in comparison to its value in $pp$ and $pA$ reactions.  
The additional suppression in this case can be explained by the  
interaction with hadronic secondaries, or ``comovers" \cite{comovers}  
(an alternative, more exotic, explanation will be discussed below).  
Indeed, the final state interactions occur at  
a later stage of the collision, when the \J\ and $\psi'$ states are  
formed. Since the $\psi'$ state has a rather large size and a tiny  
binding energy of $\simeq 50$ MeV, it can be easily destroyed  
by the interactions with hadrons. Calculations  
show \cite{GavV}, \cite{CK}, \cite{KLNS} 
that this scenario of $\psi'$ suppression in S-U collisions  
is indeed plausible. On the other hand, the \J\ suppression  
in S-U collisions is fully described by the pre-resonance nuclear  
absorption \cite{KLNS}, without any sign of an additional absorption in the final  
state. This observation lends support to the short-distance QCD  
calculations \cite{Peskin}, \cite{Kaidalov}, \cite{KS} 
that predict a very small value of \J\ dissociation cross  
section in its interactions with light hadrons at low energies.  
 
Can one describe the \J\ suppression observed in Pb-Pb collisions  
in the hadronic comover scenario?   
If one considers the \J\ dissociation cross section in its interactions  
with hadronic secondaries as a free adjustable parameter, then  
the calculations show that the {\it magnitude} of the observed suppression  
can be explained \cite{GV}, \cite{CK}, \cite{KLNS}; for cascade calculations,  
see \cite{cascade}. However, once the parameters of the calculation are  
fixed, one should be able to understand the \J\ suppression  
(or more precisely, the absence of it) in S-U collisions as well.  
This appears to be very difficult. Indeed, the atomic number dependence  
of total multiplicity produced in AB collisions at SPS energies  
is known to scale reasonably well with the number of  
participants\footnote{This concerns only the {\it total} multiplicity;  
the yield of strange particles, for example, does not follow this  
simple scaling, but this is almost irrelevant for the \J\ suppression.}. 
At first glance, this scaling looks trivial, but it is not;  
a na{\"{\i}}ve superposition of individual nucleon--nucleon collisions  
would result in the scaling with the number of {\it collisions} instead.  
The physics at work here can be understood if we again recall the existence  
of formation zone (\ref{dist}) in high energy soft processes.  
If the length of the formation zone is larger than the size of the nucleus, 
the formation of hadronic secondaries (accompanied by the energy loss 
discissed above) will take place only after  
the nuclei have already passed through each other; the multiplicity  
of produced hadrons in this case will be proportional to the number  
of inelastically excited (``wounded" \cite{Bialas}) nucleons, and not to the number  
of collisions. The density of produced secondaries in this picture  
is proportional to the density of wounded nucleons in the transverse  
plane, which can be computed using the Glauber theory. To address  
the \J\ suppression, one has also to take into account the fact that  
the \J\ distribution in the transverse plane is determined by the  
nuclear overlap function (\J\ production is a hard process with  
short formation length -- see (\ref{long})). This leads to an effective  
increase of the average density of hadronic secondaries which is ``seen"   
by \J's.    
Calculations based on this approach show  
that the average density  
of secondaries which interact with \J's increases only by $\simeq 10\%$  
from S-U to Pb-Pb system \cite{KLNS}. 
This implies a difficulty in reconciling  
the absence of additional \J\ suppression in S-U collisions with  
a strong suppression observed in Pb-Pb. One can still try to  
adjust the parameters of this model to interpolate between S-U and Pb-Pb, 
but the fit appears unacceptably poor.  
 
To overcome this problem of the hadronic model, one has to assume that the density of hadronic  
secondaries increases faster than the density of wounded nucleons from S-U to Pb-Pb.  
A calculation of this kind was performed in Ref.\cite{AC}; basing on the dual parton model,  
the authors assume that the density of hadronic  
secondaries contains two terms: a component proportional to the number of wounded  
nucleons and a component proportional to the number of collisions.  
The relative strength of the two components is an adjustable parameter of the model.  
Using this revised version of the earlier approach \cite{CK}, the authors find  
a better fit to the experimental \J\ survival probability.  
It remains to be seen, however, whether the model in its present version is   
consistent with the minimum bias and Drell-Yan--associated transverse energy spectra  
in both S-U and Pb-Pb collisions, as well as with the correlation of energy deposited  
in the transverse ($E_T$) and forward ($E_{ZDC}$) directions, measured for Pb-Pb.  
 
Irrespectively of any details of specific models based on  
final state hadonic interactions, none of them predicts a discontinuity in the  
$J/\psi/DY$ ratio -- the predicted suppression is always a smooth function  
of atomic number and centrality of the collision.  
 
\section{INTERACTION WITH PARTONIC SECONDARIES}       
  
Hard partons produced in the nucleus-nucleus collision should be very effective  
in breaking up charmonium states. The gluon--\J\ inelastic scattering 
(a ``gluo-effect''; the magnitude of the corresponding cross section was 
first estimated in ref \cite{Shuryak}),  
is expected to have an energy dependence which is very different from 
the energy dependence  
of \J\ --hadron inelastic scattering; this leads to very 
distinct absorption rates  
of \J\ in partonic and hadronic systems, and again points to the possibility  
to use \J\ and other tightly bound quarkonium states as effective probes of  
the state of QCD matter \cite{KS},\cite{KSN}. Unlike the original coherent mechanism  
of Debye screening \cite{MS}, the gluo-effect mechanism is incoherent, and requires 
only  
the presence of sufficiently hard (deconfined) gluons at the stage when the physical \J\ states  
are already formed. The relative importance of the two mechanisms is 
difficult  
to estimate at present; one needs to know in detail, in particular, 
the density dependence of the \J\ binding energy.  
 
At high energies, the nucleus-nucleus collisions are expected to 
produce a large  
number of semi--hard partons \cite{semihard}.  
These partons can then interact among themselves and  
with the produced \J's. The \J\ survival probability at RHIC and LHC energies  
in this picture was considered in ref \cite{XKSW}. The density   
of semi-hard partons is usually assumed to be proportional to the number of  
individual nucleon-nucleon collisions, since their proper formation 
time $\sim 1/P_T$  
is rather short. The \J\ survival probability therefore is a steep function  
of the atomic number of the colliding nucei and centrality of the collision.  
 
The authors of ref \cite{GM} considered an interesting possibility that 
semi-hard processes  
dominate the production of secondaries already at SPS energies. In this 
case the  
partonic density achieved in Pb-Pb collisions is much higher than in 
the S-U system;  
this allows therefore for a much stronger \J\ suppression in the former case.  
It would be interesting to check this conjecture against the available 
SPS data on  
the minimum bias and Drell-Yan associated transverse energy production, 
as well as on the centrality dependence of multiplicity.  
 
Incoherent partonic effects, as well as all other effects considered so 
far, cannot  
however produce a discontinuity in the \J\ survival probability, unless  
one assumes that something dramatic happens to \J\ {\it only} after 
the ``critical  
density'' is achieved. This brings us to the next, and most speculative, part  
of this overview. 
 
\section{...DECONFINEMENT?} 
 
The difficulties of conventional approaches outlined above have inspired   
several authors, extending the earlier model of \cite{GS}, to  
assume that once the density of produced particles exceeds some critical   
value, the formation of a ``deconfined phase'' \cite{dec}, \cite{dec1} or  
``string percolation" \cite{dec2} takes place. In practical terms,  
the survival probability of \J\ is assumed to be equal to zero if  
it is produced in the region where the density of produced particles  
exceeds some ``critical'' value. Since no anomalous \J\ 
absorption was observed in S-U  
collisions, this critical density has to exceed the maximal density  
achievable in this system. The ways in which different authors  
evaluate the density of produced particles vary somewhat, but all of them agree  
that the {\it magnitude} of \J\ suppression observed in central Pb-Pb  
collisions can be reproduced in this picture.  
      
However even this approach, aimed at introducing the most sharp discontinuity  
in the \J\ survival probability, appears to be {\it incapable} of reproducing  
the jump in the $J/\psi/DY$ ratio observed experimentally.  
The reason is easy to understand: because of the fluctuations in the  
number of produced secondaries, each value of the measured transverse  
energy $E_T$ actually corresponds to a rather broad range of the collision impact  
parameters; for the Pb-Pb system one typically finds  
an uncertainty of $1-3$ fm's. This effect leads to a {\it gradual}  
increase of \J\ suppression as a function of the measured $E_T$.  
We see that even this dramatic assumption does not lead to the explanation of   
the observed sharp discontinuity of the $J/\psi/DY$ ratio, and this is very puzzling.  
 
An interesting alternative realization of the deconfinement scenario was presented  
by {\mbox E. Shuryak} at this Conference \cite{Shuryak1}; 
in their approach, the produced deconfined   
phase reaches its ``softest point'' at some centrality in Pb-Pb collisions.  
This leads to a very long lifetime of the produced plasma, which can  
therefore effectively dissociate the produced \J's. (In this picture,  
one has to consider explicitly the finite dissociation rate of \J\ in deconfined  
matter; an estimate for this quantity can be found in ref \cite{LMS}).  
A distinctive feature of this approach is that the \J\ absorption  
is maximal at some value of centrality, corresponding to the ``softest point''  
of the equation of state of the produced deconfined phase;  
once the centrality increases further, the \J\ survival probability increases again. 
However, also in this approach, the sharp discontinuity is difficult to explain, and we  
are still left with the ``jump puzzle''. 
 
An attempt to interpret the presence of discontinuity in the $J/\psi/DY$ ratio  
was undertaken in ref \cite{KNS}. The authors were motivated by the idea  
that the formed deconfined phase should occupy some minimal volume;  
it does not make sense to consider a droplet of a new thermodynamical phase  
of a size, say, less than $1$ fm. In the nucleation theory, this size  
appears as a critical size of the bubble of a new phase in a first order phase transition.  
This minimal critical size then enters as an additional (to the critical density) parameter  
of the model. It was found that this assumption makes the description  
of the  $J/\psi/DY$ discontinuity possible (see Fig. 1)\footnote{The discontinuity 
appears as a result of dissociation of $\chi$ states at the deconfinement point; 
$\chi$'s contribute $\simeq 40\%$ 
to the \J\ production.}
 ; however the model as it  
stands at present is rather {\it ad hoc}. One may also worry about the consistency  
of the approach: indeed, the formation of equilibrated superheated hadron phase, which  
then undergoes a first order transition to the deconfined phase looks  
unlikely in a nucleus-nucleus collision. We have to keep in mind, however,  
the peculiarity of the theory we are dealing with -- the ground state of QCD,  
filled with strong color fields, is, in a way, itself a statistical system.   
A large energy density of QCD vacuum, reflected by the phenomenological  
value of the gluon condensate \cite{SVZ}, makes it a rather robust structure.    
However when the vacuum is disturbed by the multiple production of partons  
in a finite volume, its structure may change \cite{LW}, and this is the process that we are  
aiming to induce. A simple, and explicitly solvable,  
example is given by the Friedberg--Lee model \cite{FL}, desribing the interaction  
of quarks with an effective $\sigma$ field  
\be 
{\L} = {\overline{\psi}} \left( i \hat{{\partial}}  - g \sigma \right) \psi - U(\sigma); 
\ee 
The effective potential of this model on the tree level 
is the sum of $U(\sigma)$ (the $\sigma$  
self--coupling potential, sought to represent effectively the self--interactions  
of gluon fields), and the term $g \sigma \bar{\psi}\psi$, linear in the scalar quark  
density. Writing the $\sigma$ self--interaction as a polynomial, one finds 
\be           
U_{eff}(\sigma) = a \sigma^2 + b \sigma^3 + c \sigma^4 + g\ \sigma \ \bar{\psi}\psi; 
\label{effpot} 
\ee 
At vanishing quark density, 
the effective potential (\ref{effpot}), with properly chosen parameters, 
possesses two minima: a global one at $\sigma=\sigma_0$,  
mimicking the presence of the gluon condensate in QCD vacuum, and another local  
minimum at $\sigma=0$, corresponding to the vacuum of perturbation theory.  
When the density of quarks $\bar{\psi}\psi$ is small, the system  
stays at the true minimum; the quarks acquire large dynamical ``mass'' $g\sigma_0$  
and are almost excluded from the physical spectrum. However once the density  
of quarks in a finite volume becomes large, the presence of the last term in 
(\ref{effpot})   
makes the minimum at $\sigma=\sigma_0$ unstable, and it can decay to the new  
true vacuum with $\sigma=0$; the decay proceeds via a formation of a finite size  
bubble of a new phase \cite{bubble}. This schematic 
model illistrates the physical phenomena  
which might occur in a nucleus-nucleus collision when the density of produced  
partons exceeds some threshold value. Whether this decay of  
QCD vacuum can occur in nucleus-nucleus collisions, and whether it is  
responsible for the observed discontinuity of \J\ survival probability  
remains an intriguing question, which still has to be answered.       
 
At the same time the behavior of $\psi'$ does not  
show any unusual behavior -- the measured $\psi'$ survival probability  
is a smooth function of $E_T$. Is it consistent with the 
existence of a threshold phenomenon?  
The answer is the following: $\psi'$'s have a large radius and a small binding energy  
and are easily dissociated by hadronic secondaries; their density is the highest  
in the central region of the transverse plane, where the number of colliding nucleons  
is the largest. Calculations show that in this region almost all of the $\psi'$'s  
are absorbed. Introducing additional suppression, for example, by the formation  
of a bubble of deconfined phase, therefore does not affect the overall survival  
probability -- the only observed $\psi'$'s are produced in the peripheral region  
of the transverse plane. In other words, the $\psi'$ suppression in Pb-Pb and 
S-U collisions can be caused both by interactions with hadronic secondaries and 
by deconfinement, and there is no easy way to distinguish between the two effects.         
          
To summarize this Section: the deconfinement scenario can accommodate the observed  
features of ``anomalous'' \J\ suppression, but only at the expense of introducing  
some model--dependent assumptions. A detailed, consistent and convincing approach  
based on the deconfinement scenario still has to be developed. However this is the only picture  
known at present that is capable of explaining the observed stunning features of the data, 
and it has to be seriously examined and explored.   
   
\section{WHAT ELSE DO WE NEED TO KNOW?} 
  
One has to admit that the problems that the theorists are facing in the physics of  
relativistic heavy ion collisions are too difficult for them to solve.  
To prove this, let me remind you that none of the theorists {\it predicted}  
the onset of anomalous \J\ suppression in Pb-Pb collisions, let alone  
the centrality at which it should begin. The advocates of deconfinement scenario,  
who made a generic prediction that the anomalous \J\ suppression should  
show up once the density is ``high enough'', at least have an excuse --  
for them, this is the phenomenon that was never observed before, and the  
behavior of QCD matter in these conditions is largely unknown.  
However, also the theorists advocating conventional explanations could  
not anticipate the onset of a stronger \J\ suppression; in this case,  
since conventional mechanisms, by definition, are supposed to be well--known, 
one should have  
been able to make a prediction. The fact that none of these predictions  
were made {\it before} the experimental discovery of anomalous \J\ suppression, 
tells us once again that the field of relativistic heavy ion collisions  
is, and most likely will remain to be, experiment--driven, and we will have to  
rely on the experimental results to make any progress.  
 
What data do we need to clarify the origin of the anomalous \J\ suppression?  
The most important thing now is to establish firmly (or to discard) the presence  
of discontinuity in the $J/\psi/DY$ ratio. A further increase in statistics, 
especially for the high mass Drell--Yan events, would be beneficial for this. 
We should also verify that a decrease of the energy of the Pb beam and/or 
the use of a lighter target lead to the disappearance of the anomalous suppression.  
This would prove the threshold nature of the phenomenon responsible for the observed effect.  
We would then know for sure that the collective behavior in QCD matter has been discovered,  
and we have many years ahead of trying to understand it.     
 
\vskip0.3cm 
   
\vskip0.2cm  
\noindent I would like to thank the Organizers of this Conference for their kind 
invitation to this most stimulating meeting.

I thank H. Satz for introducing me into this fascinating problem, and for 
our continuing collaboration. I am indebted to those of my colleagues who have 
tried to teach me the topics reviewed here; I am particularly thankful to 
J.-P. Blaizot, A. Capella, \mbox{Yu.L. Dokshitzer}, \mbox{K.J. Eskola}, 
\mbox{E.L. Feinberg}, S. Gavin, 
K. Geiger, C. Gerschel, 
\mbox{M. Gonin,}
M. Gyulassy, \mbox{J. H\"{u}fner}, \mbox{F. Karsch}, L. Kluberg, 
B.Z. Kopeliovich, 
C. Louren\c{c}o, 
\mbox{T. Matsui,} \mbox{L.McLerran,} 
\mbox{A.H.Mueller,} B.M\"{u}ller,  \mbox{M.Nardi},   
J.-Y.Ollitrault, Y.Pang, J.Qiu, \mbox{L.Ramello}, \mbox{J. Schukraft}, 
E.V. Shuryak, \mbox{X.-N. Wang}, X.-M. Xu,    
G.M. Zinoviev and \mbox{R. Vogt}.

\end{document}